\documentstyle[prd,aps,epsfig,floats]{revtex}
\begin{document}

\draft
\renewcommand{\topfraction}{0.8} 
\newcommand{\beq}{\begin{equation}}
\newcommand{\eeq}{\end{equation}}
\newcommand{\bea}{\begin{eqnarray}}
\newcommand{\eea}{\end{eqnarray}}
\newcommand{\pbar}{\not{\!\partial}}
\newcommand{\dbar}{\not{\!{\!D}}}
\def\lsim{\:\raisebox{-0.75ex}{$\stackrel{\textstyle<}{\sim}$}\:}
\def\gsim{\:\raisebox{-0.75ex}{$\stackrel{\textstyle>}{\sim}$}\:}
\twocolumn[\hsize\textwidth\columnwidth\hsize\csname 
@twocolumnfalse\endcsname

\title{Enhanced Reheating via Bose Condensates} 
  
\author{Rouzbeh Allahverdi$^{a}$, Robert Brandenberger$^{b,c,d}$ and Anupam 
Mazumdar$^{d}$}

\address{$^a$ Theory Group, TRIUMF, 4004 Wesbrook Mall, Vancouver, BC, V6T
2A3, Canada. \\
$^b$ Physics Dept., Brown University, Providence, RI 02912, USA.\\
$^c$ Perimeter Institute, Waterloo, ON N2J 2W9, Canada.\\
$^d$ Physics Dept., McGill University, 3600 University Road, Montr\'eal, QC,
H3A 2T8, Canada.}
\date{\today} 
\maketitle

\begin{abstract}

In supersymmetric extensions of the particle physics Standard Model, 
gauge invariant combinations of squarks and sleptons (flat directions) 
can acquire large expectation values during a period of 
cosmological inflation. If the inflaton sector couples to matter fields via 
these flat directions, then new channels for efficient reheating,
in particular via parametric resonance instabilities, are opened up.
These can lead to efficient reheating induced by the flat directions even if
the bare coupling constants are small. In this Letter we discuss various
channels which yield this ``enhanced reheating'' effect, and we address some 
cosmological consequences.

\end{abstract}


\vskip2pc]


\section{Introduction}

There are strong indications that the early Universe might have
undergone a phase of inflationary expansion~\cite{WMAP}. The success of big
bang cosmology puts severe constraints on early Universe
cosmology, and therefore, if inflation is the right paradigm, then it must
yield a way of understanding the origin of all observed matter, including 
the baryons of the Standard Model (SM). In spite of many attempts, it has proven
hard to pin down the inflaton sector. Since its couplings with standard
model fields in the bare Lagrangian are constrained to be very small,
it is usually assumed that the inflaton is a 
SM gauge singlet, and in most models it is even
considered as an absolute gauge singlet (for reviews see e.g. 
\cite{Lindebook,Lyth}). 
This leads to the question of how the Universe
reheats and produces the matter we observe today.

It is commonly believed that the inflaton sector couples to either the SM
sector or the matter sector of the minimal extension of the SM, 
the Minimal Supersymmetric Standard Model (MSSM) through
non-renormalizable interactions (for a review see~\cite{Nilles}). 
In this context, however, it is not clear why the inflaton would decay at
a sufficient rate into the SM degrees of freedom. This is the issue
we will focus on in this Letter. We will concentrate on the case when
the inflaton couples to either SM or MSSM fields via
interactions which are suppressed by the scale at which new
physics is being invoked, which could be either the GUT scale, string
scale or the Planck scale.
It turns out that the careful consideration of the evolution of the
fields corresponding to the flat directions leads to a new mechanism of
reheating which we call ``enhanced reheating'' 

As is well known, MSSM has flat directions, made up of gauge invariant
combinations of squarks and sleptons (possibly including right-handed
sneutrinos), which may acquire non-vanishing
expectation values during inflation (see e.g. \cite{Lindebook} for a review), 
thereby forming homogeneous condensates. The condensates may play a
significant role in many cosmological phenomena (see e.g. \cite{Kari} for
recent reviews), such as generating the baryon asymmetry, or 
producing the dark matter particles 
by first fragmenting into $Q$-balls which then decay \cite{qball}. It 
has also been suggested that the cosmological density perturbations
could be due to fluctuations of a
MSSM flat direction condensate \cite{denspert}. In this Letter we will
study novel effects of MSSM flat direction condensates for reheating.


\section{Enhanced Perturbative Reheating}

Even if at the level of the bare particle physics Lagrangian there is
no coupling between the inflaton and the MSSM matter fields, such
a coupling will be generated via gravitational effects. The
resulting terms will typically correspond to a non-renormalizable
field theory and will be suppressed by the Planck mass.
We will model these terms via a simple superpotential term
\beq \label{superpot}
W \supset \lambda {\Phi \over M_{\rm P}} {\bf X} {\Psi}_1 {\Psi}_2 \, ,
\eeq
where $\Phi$ is the (singlet) inflaton superfield, ${\bf X}$ is the superfield 
containing
the condensate, $\Psi_i$ stands for MSSM matter superfields, ${M}_{\rm P} = 
2.4 \times 10^{18}$ GeV is the reduced Planck mass, and $\lambda$ is a 
coefficient. Note that all what is needed 
is that ${\bf X} \Psi \Psi$ be a gauge 
invariant combination of the MSSM fields. Then, regardless of any discrete or 
continuous global symmetry (e.g. $R$ symmetry) that the theory may have, 
gravitational effects are expected to generate the above superpotential 
term. If the inflaton is a gauge singlet up to $M_{\rm P}$, $\lambda \sim 
{\cal O}(1)$ typically. On the other hand, if $\Phi$ is non-singlet under new 
physics at a scale $M_{\rm new} < M_{\rm P}$, $\lambda$ will generally be 
suppressed by powers of $M_{\rm new}/M_{\rm P}$.    

Around the minimum of its potential, the inflaton superpotential can be
approximated by $m_\phi {\Phi}^2/2$. At the end of inflation $\langle
X \rangle = X_I$ and, treating the inflaton as a real scalar field,
eq.~(\ref{superpot}) results in a Lagrangian of the form
\beq \label{lagr}
{1 \over {\lambda}} \left({M_{\rm P} \over X_I}\right) {\cal L} 
\supset \phi {\bar \psi} \psi
+ m_\phi \phi ({\chi}^2 + {{\chi}^{*}}^{2}) 
+ \lambda \left({X_{\rm I} \over M_{\rm P}}\right) \phi^2 {\vert {\chi} 
\vert}^2,
\eeq
where $\psi$ is a Dirac fermion containing the 
fermionic components of ${\Psi}_1$ and ${\Psi}_2$, and ${\chi}$ 
denotes the scalar components of the $\Psi$ superfields.
 
A phase of primordial inflation driven by the slowly rolling scalar
field $\phi$ sweeps the Universe clean of any particle excitations which might
have been present before inflation, 
but leads to vacuum fluctuations which generically will
produce quasi-homogeneous expectation values (condensates) of light fields
such as $X_I$. After the slow-roll approximation breaks down, the
inflaton field will begin a phase of coherent oscillations about the minimum
of its potential. During this phase, the coherent oscillations will
dominate the energy density of the Universe. Due to couplings of the inflaton
field to other fields, the oscillations will transfer some of
their energy density to the matter particles. Upon 
thermalization of these particle excitations, a radiation-dominated
Friedmann-Robertson-Walker (FRW) universe will result, as in the hot big-bang
model. This process is called {\it reheating}. Successful Big Bang
Nucleosynthesis (BBN)~\cite{bbn} requires that the reheat temperature
of the universe obey $T_{\rm R} \gsim {\cal O}(\rm MeV)$. 

In the initial analyses \cite{reheat}, 
the energy transfer from the coherent inflaton
field to matter was treated perturbatively, using the lowest order
analysis to determine the rate $\Gamma_{\rm d}$ with which
a single $\phi$ particle at rest decays into quanta of the fields
which $\phi$ couple to. While the Hubble constant $H$ exceeds the decay
rate, the inflaton will lose energy primarily to the expansion of space,
but as soon as $H$ drops $\Gamma_{\rm d}$, the energy of $\phi$ goes
primarily into matter particles, resulting in a gas of matter with
a temperature $T_{\rm R}$ given by 
\beq \label{reheattemp}
T_{\rm R} \, \sim \, 0.1 {({\Gamma}_{\rm d}{M}_{\rm P})}^{1/2} \, .
\eeq

In our toy model (\ref{lagr}), the inflaton can decay perturbatively via
several channels. In the limit that the inflaton mass $m_{\phi}$ is much
larger than the mass of $\psi$ and $\chi$, the perturbative decay rate
of $\phi$ into fermion  and scalar pairs is given by
\beq \label{xdecay}
\Gamma_{I} \, = \,  
\lambda^2 {m_\phi \over 4 \pi} \left({X_{\rm I} \over M_{\rm P}}\right)^2 \, .
\eeq     
Let us assume that the perturbative decay rate in the absence of the $X$
condensate is $\Gamma_0$ (with associated reheat temperature $T_{\rm R, 0}$
determined via (\ref{reheattemp})).
From (\ref{xdecay}) it immediately follows that a large value of
$X_I$ leads to {\it enhanced perturbative reheating}.
 The condition for enhanced reheating 
(i.e. $\Gamma_{\rm I} > \Gamma_{0}$) is
\beq \label{dom}
\lambda^2 \left({X_{\rm I} \over M_{\rm P}}\right)^2 \, > \, 
10^3 {T^2_{\rm R,0} \over m_\phi M_{\rm P}}.
\eeq
As an example, consider the chaotic inflation model with $m_\phi = 10^{13}$ 
GeV and $T_{\rm R,0} = 10^9$ GeV. Then the bound in~(\ref{dom}) translates to 
$\lambda X_{\rm I} > 10^{13}$ GeV.


\section{Enhanced Parametric Reheating}

As first discussed in \cite{tb} (see also \cite{DK}), in 
many models there is a non-perturbative and much more efficient
reheating mechanism which makes use of the parametric resonance
instability in the field equation of the particles being produced
($\psi$ and $\chi$ in our model) in the presence of an oscillating
inflaton field. Note that, as first discussed heuristically in
\cite{tb} and then analytically in \cite{KLS1} and \cite{STB}, the
instability is present even when taking the expansion of space and
the resulting decay of the inflaton oscillation amplitude into
account. This mechanism leads to an initially highly non-thermal
distribution of particles (including inflatons), while full thermal 
equilibrium takes much longer to establish (see e.g. \cite{fk} for
a recent study).

The Lagrangian (\ref{lagr}) contains three separate channels for
resonance. The last term on the right-hand side of (\ref{lagr}) 
corresponds to two-particle induced resonant production of
$\chi$ bosons and was treated in detail in \cite{KLS2}. 
The second term leads to single-particle induced resonant production 
of $\chi$ bosons and was analyzed in \cite{STB}, while the first
term lead to parametric production of $\psi$ fermions and was
treated in detail in \cite{GK}. The third term dominates for large
values of $X_I$.

Let us first focus on the third term on the right hand side of (\ref{lagr}).
If we neglect the other two terms, then in the presence of the oscillating
inflaton field, the equation of motion for $\chi$ will lead to
exponentially growing solutions with a Floquet exponent $\mu$ much larger
than the Hubble expansion rate. The instability occurs for a large
range of comoving momenta (and thus the instability is called of
{\it broad resonance type} \cite{KLS2}). The instability is
effective as long as \cite{KLS2}
\beq \label{brcond}
\lambda \bigl({{X_I} \over {M_{\rm P}}}\bigr) \phi \, > \, m_{\phi} \, .
\eeq

The effective coupling constant during inflation is bounded by the
amplitude of the induced CMB anisotropies, leading to the constraint
\beq \label{CMBconstraint}
\lambda^2 \bigl({{X_I} \over {M_{\rm P}}}\bigr)^2 \, \leq \, 10^{-6} \, , 
\eeq
making use of the fact that an interaction coupling constant arising in
(\ref{lagr}) will induce a one-loop correction to the quartic inflaton
self coupling which is bounded by the standard fluctuation analysis (see
e.g. \cite{MFB} for a review).

During inflation, long wavelength quantum fluctuations of $X$ 
accumulate in a coherent state with a maximum
quasi-homogeneous field value given by \cite{Lyth}
$\langle X^2 \rangle = 3 H^4_{\rm I}/8 \pi^2 m^2_X$, which for
condensate masses smaller than the Hubble constant during inflation
can be many orders of magnitude larger than $H_{\rm I}$.  The exception is when
supersymmetry breaking by the inflaton energy density~\cite{drt}
results in a correction $\approx + H^2_{\rm I}$ to the $(\rm mass)^2$
of $X$ during inflation.\footnote{Note that a correction $\approx - H^2$ 
typically results in $X_I \gg H_I$, since $X$ will in this case settle at 
the minimum of the potential which is far away from the origin.} Another 
way to obtain a large expectation value $X_I$ is if $X$ is exactly
massless during inflation and obtains a non-vanishing mass during
a phase transition which takes place during the early stages of
reheating, in analogy to how the QCD axion field develops an 
expectation value after the QCD phase transition. 
Note that for field values larger than
$M_{\rm P}$, gravitational corrections to the potential may not
be under control, and hence we will consider 
$H_{\rm I} \ll X_{\rm I} \leq M_{\rm P}$ at the end of inflation. 

The scenario of {\it enhanced parametric resonance} is now as follows. At
the end of the slow-rolling phase of inflation, $\phi \sim M_{\rm P}$, and, 
since the inflaton mass $m_{\phi}$ is
constrained by the amplitude of the CMB fluctuations to satisfy
$m_{\phi} < 10^{-6} M_{\rm P}$, the broad resonance condition (\ref{brcond})
is satisfied and resonant $\chi$ production occurs. As a
consequence of the energy loss to particle production and of the expansion
of space, the amplitude of $\phi$ oscillations will decrease. If the 
condition (\ref{brcond}) is satisfied when resonant particle production 
becomes explosive, inflaton oscillations will decay in the 
broad resonance regime~\cite{KLS2}. When the condition (\ref{brcond}) 
ceases to be satisfied the broad resonant decay shuts off. 

Even after the period of broad resonance ends,
parametric resonance via the first two terms in
(\ref{lagr}) will continue for a while. To see this, focus on the
resonant production of $\chi$ particles via the single $\phi$ coupling
in (\ref{lagr}). The perturbative decay rate is the same as the
perturbative decay rate (\ref{xdecay}) into fermions via the first term on
the right hand side of (\ref{lagr}). As studied in detail in \cite{STB}, the
parametric resonant decay of $\phi$ proceeds by exciting particles
in narrow resonance bands, the first at $k = m_{\phi}/2$. Decay into
the lowest resonance band dominates the overall decay rate, and, focusing
on this band, the decay rate is given by
\beq
\Gamma_{\chi} \, \sim \, {{H m_{\phi}^2} \over {16 \pi^2 \phi^2}}
sinh^2 \bigl[\lambda^2 ({{X_I} \over {M_{\rm P}}})^2 {{\phi^2} \over 
{H m_{\phi}}} \bigr] 
\eeq
(see Eq. (125) of \cite{STB}).

If the argument of the hyperbolic function is smaller than 1, the result
reduces to the perturbative decay rate. Thus, this channel has enhanced
parametric reheating provided the argument is larger than 1, which is the
case if (taking into account the fact that the inflaton potential is taken
to be quadratic)
\beq \label{effcond}
\lambda^2 {{X_I^2 \phi} \over {M_{\rm P} m_{\phi}^2}} \, > \, 1 \, .
\eeq
By inserting the saturation values of (\ref{brcond}) and
(\ref{CMBconstraint}) 
into (\ref{effcond}), it follows immediately that this condition remains 
satisfied once the broad resonant decay shuts off, provided that
$M_{\rm P} > 10^3 m_{\phi}$. 

Once the value of $H$ drops below $m_X$, the $X$ condensate will also begin 
damped oscillations.
Thus, the conditions in (\ref{brcond}), respectively (\ref{effcond}), may 
cease to be 
satisfied before broad, respectively narrow, resonant decay completes.
However, the oscillations of $X$ lead to another channel of parametric
reheating, namely reheating induced by the coupling of $X$ to other
matter fields in the part of the Lagrangian which is independent of the
inflaton (see \cite{Postma} for a discussion of this scenario).

To conclude this section, we have demonstrated that large expectation values
of scalar condensates such as arise in the MSSM open up several channels
for enhanced reheating via parametric resonance.


\section{Some Consequences}

Let us first discuss the consequences of the above analysis for the 
``temperature''
$T_R$ after reheating, by which we mean the fourth root of the energy
density in matter after the inflaton has lost a majority of its energy
into matter \footnote{If the state after this stage of ``preheating'' 
\cite{KLS1}
were thermal, then the quantity $T_R$ thus computed would indeed be the
usual thermodynamical temperature of matter. However, if the reheating
proceeds via parametric resonance, then the state of matter immediately
after preheating will be far from thermal.} There are various possible
scenarios.

In the first case, the broad resonance condition (\ref{brcond}) is satisfied
when the inflaton begins to oscillate, which occurs when $\phi \sim M_{\rm P}$.
In this case, the energy transfer into matter happens explosively via broad
parametric resonance, yielding the value (remembering again that the
potential for $\phi$ is quadratic)
\beq \label{rhtemp}
T_R \, \sim \, \bigl( m_{\phi} M_{\rm P} \bigr)^{1/2} \, ,
\eeq
which, for values $m_{\phi} \sim 10^{-6} M_{\rm P}$ obtained from the 
amplitude of the observed CMB spectrum under the hypothesis that
adiabatic perturbations of the inflaton generate the observed anisotropies,
is about $10^{16}$ GeV. This scenario is realized quite naturally if
the initial value of the condensate is close to the Planck scale.

If the broad resonance condition (\ref{brcond}) is not satisfied when the
inflaton starts oscillating, or ceases to be satisfied before broad resonant 
decay completes, but the condition (\ref{effcond}) for
effectiveness of the narrow resonance instability via the second interaction
term on the right hand side of (\ref{lagr}) is satisfied, then once again
the energy transfer to matter will be immediate and the ``temperature'' after
the preheating will be given by (\ref{rhtemp}). 

The third scenario arises when neither (\ref{brcond}) nor (\ref{effcond}) are
satisfied, but when the perturbative decay rate in the presence of the
condensate $\Gamma_{\rm I}$ is larger than  $m_X$, the mass of the condensate.
In this case, reheating completes before $X$ starts its oscillations. In this
case, the reheat temperature is given by (\ref{reheattemp}), 
with $\Gamma_{\rm d} = \Gamma_I$, yielding 
\beq \label{trenh}
T_{\rm R} \, \simeq \, 3 \times 10^{-2} \lambda X_{\rm I} \left({m_{\phi} 
\over M_{\rm P}}\right)^{1/2} \, .
\eeq

Finally, if none of the above resonance conditions are satisfied and in
addition $\Gamma_{\rm I} < m_X$, then the condensate will begin to
oscillate once the Hubble parameter $H$ drops below $m_X$, and this
will lead to new production channels for the production of matter via
$X$ decay. Assuming that the decay is rapid on a Hubble time scale, 
the resulting reheat ``temperature'' $T_R$ (quotation makes indicating
once again that the quantity is simply a measure of the energy density of the
matter particles after the decay of the $X$ condensate) will be given by
\beq
T_{\rm R} \, \sim \, \bigl( m_X X_I \bigr)^{1/2} \, ,
\eeq
unless the result is smaller than what is obtained from (\ref{trenh}), in
which case most of the energy of matter particles results from the inflaton
decay which starts once $\Gamma_{\rm I} = H$, i.e. after the condensate has
started to oscillate. 

We have seen that reheating can be efficient even for very
small inflaton couplings to matter. This can have good or bad
effects for early Universe cosmology. On the positive side,
the enhancement of reheating by the $X$ condensate may make it possible
to obtain reheating temperatures sufficiently high for
successful baryogenesis and/or dark matter production even when the
reheating temperature $T_{\rm R,0}$ (calculated using the perturbative
decay rate in the absence of the condensate) is too low.  
However, enhanced reheating should not lead to the
gravitino problem~\cite{gravitino}. If $T_R > 10^9$ GeV, then
there will be a gravitino overproduction problem unless
the gravitino mass $m_{3/2}$ obeys  $m_{3/2} \geq 20$ TeV 
(this happens, for example, in models of anomaly-mediated
supersymmetry breaking~\cite{anomal}). Otherwise, late entropy
generation will be required to dilute the excess of gravitinos.
As we have seen above, if the parametric resonance channels for
reheating are open as a consequence of a large expectation value of
$X$, then $T_R$ can be of the order $10^{16}$ GeV and we will be
faced with the gravitino problem. Even if the parametric resonance
channels are closed, then, using (\ref{trenh}), we conclude that
there will be a gravitino problem if
\beq
\lambda X_I \, > \, \, 10^{-5} M_{\rm P} \, .
\eeq

A second stage of inflaton like in {\it thermal inflation} \cite{thermalinfl}
can dilute the gravitino density and thus solve the gravitino
problem. If $X_I \geq M_{\rm P}$ at the end of inflation, then
such a second stage of inflation might naturally arise in our
framework: once $H$ drops below the value of $m_X$ and the condensate
can start to move, it will initially be slowly rolling and
thus generate a second period of inflation.

Another consequence of the enhanced reheating scenario discussed here
concerns the origin of cosmological fluctuations. As is well known
quantum fluctuations of the $X$ condensate will lead
to isocurvature fluctuations as in the case of axions (see e.g. \cite{axion})
which upon the decay of the condensate will convert to adiabatic
fluctuations via the curvaton mechanism \cite{curvaton}. However, the
$X$ fluctuations will also lead to a spatial modulation of the reheating
rate on super-Hubble scales and will thus generate fluctuations
via the {\it modulated decay} mechanism \cite{modulated} (this has
already been pointed out in the context of the MSSM in \cite{anupam}. 
As discussed
in \cite{FB}, the fluctuations which are generated in this way
are primordial isocurvature fluctuations, and their production is
completely consistent with causality.

The situation will be different when 
\beq \label{slow}
\Gamma \, < \, m_X \, ,
\eeq
$X_I < M_{\rm P}$, and if the parametric resonance
channels for enhanced reheating are not open. In this case, the $X$
field starts oscillating at $H = m_X$, before reheating
completes. From then on, $\langle X \rangle \propto H$ leading to
$\Gamma_I(t) \propto H^2$. This implies that reheating will
remain inefficient until $\Gamma_{\rm d}$ settles at its value in the
minimum $\Gamma_{0}$. The reheat temperature of the universe will then
be $T_{\rm R,0}$. 

An important issue to take into consideration is early oscillations of $X$ 
due to thermal effects from perturbative reheating~\cite{thermal}. The $X$ 
field has gauge
and/or Yukawa couplings (collectively denoted by $h$) to other fields.
If $h X_{\rm I} \leq T$, these degrees of freedom will be in
equilibrium with the instantaneous thermal bath from reheating. This in turn 
induces
a thermal mass $hT$ for the $X$ field which will trigger its early
oscillations if $hT \geq H$. Early oscillations, if they start before 
reheating completes, would render the reheating inefficient
as noted above. There will be no early oscillations if $h X_I > T$ or $hT < 
H$. This requires sufficiently large or small values of $h$, respectively, 
which can naturally occur for the MSSM flat directions and right-handed 
sneutrinos respectively. The early oscillations will not be important in the 
case of parametric reheating. In this case, 
fields which are coupled to $X$ via 
gauge and/or Yukawa couplings reach thermal equilibrium on time scales much 
longer than that of the resonant inflaton decay~\cite{fk}.


\section{Conclusions}

In this Letter we have discussed {\it enhanced reheating channels}
which arise as a consequence of the presence of scalar condensate
fields which acquire a large expectation value $X_I$ during inflation. The
scalar condensates can result in large effective couplings of the inflaton
to matter, which can enhance both the perturbative
decay rate and the efficiency of parametric resonance instabilities.
We have seen that
efficient reheating is therefore possible even if inflaton has only
gravitational couplings to matter fields in the bare Lagrangian.
 
Enhanced reheating can have many consequences for cosmology. Generically,
the reheating temperature will be higher than what would result in the
absence of the condensates. In supersymmetric models this can make
the gravitino problem worse.

Since scalar condensates acquire fluctuations during inflation
which are super-Hubble-scale at the end of inflation, they provide
a realization of the modulated fluctuation scenario of \cite{modulated}:
at the time of reheating, super-Hubble scale entropy perturbations
are generated as a consequence of the space-dependent effective
coupling constants. This will relax the bounds from density perturbations on 
the inflaton mass $m_\phi$.

In summary, scalar condensates can lead to a successful scenario of chaotic 
inflation where the inflaton has only gravitational couplings and $m_\phi \ll 
10^{13}$ GeV. Immediately after the phase of slow-roll inflation comes to
an end and the inflaton field begins to oscillate, the large amplitude
of the condensate $X_I$ leads to large effective couplings in the 
dimension 5 and 6 operators which couple the inflaton, the condensate and 
matter fields. This opens up channels for decay of the inflaton via
broad parametric resonance which would be closed in the absence of $X$. Thus,
rapid reheating takes place. The decay will produce massive particles
which can provide the dark matter and/or (in the case of right-handed
neutrinos) the lepton asymmetry required for leptogenesis.
At the same time, the super-Hubble-scale
fluctuations of the condensate lead to isocurvature fluctuations both of 
axionic type and of the type generated through modulated decay.    
A detailed study of enhanced
reheating and its consequences for baryogenesis and dark matter will
appear in a separate publication.

\section*{Acknowledgments}

The research of RA and AM is supported by the National Sciences and
Engineering Research Council of Canada. AM is a CITA National Fellow.
RB is supported in part by the US Department of Energy under Contract
DE-FG02-91ER40688, TASK~A. He thanks the Perimeter Institute for their
gracious hospitality and financial support during the course of the
work on this project. AM thanks the ASPEN center for physics for their
kind hospitality during the course of this work.



\begin{references}
%
\bibitem{WMAP}
C. L. Bennett. {\it et.al.}, 
Astrophys. J. Suppl. {\bf 148},1 (2003).
%
\bibitem{Lindebook} 
A. D. Linde, {\it Particle Physics and Inflationary Cosmology},
Harwood, Chur, Switzerland (1990).
%
\bibitem{Lyth}
D.~H.~Lyth and A.~Riotto,
Phys.\ Rept.\  {\bf 314}, 1 (1999)
[arXiv:hep-ph/9807278].
%
\bibitem{Nilles}
H.~P.~Nilles,
Phys.\ Rept.\  {\bf 110}, 1 (1984).
%
\bibitem{Kari}
K.~Enqvist and A.~Mazumdar,
Phys.\ Rept.\  {\bf 380}, 99 (2003)
[arXiv:hep-ph/0209244];\\
M.~Dine and A.~Kusenko,
Rev.\ Mod.\ Phys.\  {\bf 76}, 1 (2004)
[arXiv:hep-ph/0303065].
%
\bibitem{qball}
A. Kusenko, and M. E. Shaposhnikov, Phys. Lett. B {\bf 418}, 46 (1998)
[arXiv:hep-ph/9709492];\\ 
G. R. Dvali, A. Kusenko, and M. E. Shaposhnikov, 
Phys. Lett. B {\bf 417}, 99 (1998)
[arXiv:hep-ph/9707423];\\ 
K.~Enqvist and J.~McDonald, Phys. Lett. B {\bf 425}, 309 (1998)
[arXiv:hep-ph/9711514];\\  
K.~Enqvist and J.~McDonald,
Nucl.\ Phys.\ B {\bf 538}, 321 (1999)
[arXiv:hep-ph/9803380].
%
\bibitem{denspert}
K.~Enqvist, S.~Kasuya and A.~Mazumdar,
Phys.\ Rev.\ Lett.\  {\bf 90}, 091302 (2003)
[arXiv:hep-ph/0211147];\\
K.~Enqvist, A.~Jokinen, S.~Kasuya and A.~Mazumdar,
Phys.\ Rev.\ D {\bf 68}, 103507 (2003)
[arXiv:hep-ph/0303165];\\
K.~Enqvist, S.~Kasuya and A.~Mazumdar,
arXiv:hep-ph/0311224 (to appear in Phys. Rev. Lett.);\\
K.~Enqvist, A.~Mazumdar and A.~Perez-Lorenzana,
arXiv:hep-th/0403044.
%
\bibitem{bbn}
K.~A.~Olive, G.~Steigman and T.~P.~Walker,
Phys.\ Rept.\  {\bf 333}, 389 (2000)
[arXiv:astro-ph/9905320].
%
\bibitem{reheat}
A.~D.~Dolgov and A.~D.~Linde,
Phys.\ Lett.\ B {\bf 116}, 329 (1982);\\
L.~F.~Abbott, E.~Farhi and M.~B.~Wise,
Phys.\ Lett.\ B {\bf 117}, 29 (1982).
%
\bibitem{tb} 
J.~H.~Traschen and R.~H.~Brandenberger,
Phys.\ Rev.\ D {\bf 42}, 2491 (1990).
%
\bibitem{DK}
A.~D.~Dolgov and D.~P.~Kirilova,
Sov.\ J.\ Nucl.\ Phys.\  {\bf 51}, 172 (1990)
[Yad.\ Fiz.\  {\bf 51}, 273 (1990)].
%
\bibitem{KLS1}
L.~Kofman, A.~D.~Linde and A.~A.~Starobinsky,
Phys.\ Rev.\ Lett.\  {\bf 73}, 3195 (1994)
[arXiv:hep-th/9405187].
%
\bibitem{STB}
Y.~Shtanov, J.~H.~Traschen and R.~H.~Brandenberger,
Phys.\ Rev.\ D {\bf 51}, 5438 (1995)
[arXiv:hep-ph/9407247].
%
\bibitem{fk}
G.~N.~Felder and L.~Kofman,
Phys.\ Rev.\ D {\bf 63}, 103503 (2001)
[arXiv:hep-ph/0011160].
%
\bibitem{KLS2}
L.~Kofman, A.~D.~Linde and A.~A.~Starobinsky,
Phys.\ Rev.\ D {\bf 56}, 3258 (1997)
[arXiv:hep-ph/9704452].
%
\bibitem{GK}
P.~B.~Greene and L.~Kofman,
Phys.\ Lett.\ B {\bf 448}, 6 (1999)
[arXiv:hep-ph/9807339].
%
\bibitem{MFB}
V.~F.~Mukhanov, H.~A.~Feldman and R.~H.~Brandenberger,
Phys.\ Rept.\  {\bf 215}, 203 (1992).
%
\bibitem{drt}
M.~Dine, L.~Randall and S.~Thomas,
Phys.\ Rev.\ Lett.\  {\bf 75}, 398 (1995)
[arXiv:hep-ph/9503303].
%
\bibitem{Postma}
M.~Postma and A.~Mazumdar,
J. Cosmol. Astropart. Phys. {\bf 0401}, 005 (2004)
[arXiv:hep-ph/0304246].
%
\bibitem{gravitino}
M.~Y.~Khlopov and A.~D.~Linde,
Phys.\ Lett.\ B {\bf 138}, 265 (1984);\\
J.~R.~Ellis, J.~E.~Kim and D.~V.~Nanopoulos,
Phys.\ Lett.\ B {\bf 145}, 181 (1984);\\
J.~R.~Ellis, D.~V.~Nanopoulos, K.~A.~Olive and S.~J.~Rey,
Astropart.\ Phys.\  {\bf 4}, 371 (1996)
[arXiv:hep-ph/9505438];\\
for discussions of non-perturbative decay see: 
A.~L.~Maroto and A.~Mazumdar,
Phys.\ Rev.\ Lett.\  {\bf 84}, 1655 (2000)
[arXiv:hep-ph/9904206];\\
R.~Kallosh, L.~Kofman, A.~D.~Linde and A.~Van Proeyen,
Phys.\ Rev.\ D {\bf 61}, 103503 (2000)
[arXiv:hep-th/9907124].
%
\bibitem{anomal}
G.~F.~Giudice, M.~A.~Luty, H.~Murayama and R.~Rattazzi,
JHEP {\bf 9812}, 027 (1998)
[arXiv:hep-ph/9810442].
%
\bibitem{thermalinfl}
D.~H.~Lyth and E.~D.~Stewart,
Phys.\ Rev.\ D {\bf 53}, 1784 (1996)
[arXiv:hep-ph/9510204].
%
\bibitem{axion}
M.~Axenides, R.~H.~Brandenberger and M.~S.~Turner,
Phys.\ Lett.\ B {\bf 126}, 178 (1983).
%
\bibitem{curvaton}
D.~H.~Lyth and D.~Wands,
Phys.\ Lett.\ B {\bf 524}, 5 (2002)
[arXiv:hep-ph/0110002].
%
\bibitem{modulated}
G.~Dvali, A.~Gruzinov and M.~Zaldarriaga,
Phys.\ Rev.\ D {\bf 69}, 023505 (2004)
[arXiv:astro-ph/0303591];\\
L.~Kofman,
arXiv:astro-ph/0303614;\\
F.~Vernizzi,
Phys.\ Rev.\ D {\bf 69}, 083526 (2004)
[arXiv:astro-ph/0311167].
%
\bibitem{anupam}
K. Enqvist, A. Mazumdar and M. Postma, Phys. Rev. D {\bf 67}, 121303 (2003)
[arXiv:astro-ph/0304187];\\ 
R. Allahverdi, astro-ph/0403351 (to appear in Phys. Rev. D).
%
\bibitem{FB}
F.~Finelli and R.~H.~Brandenberger,
Phys.\ Rev.\ Lett.\  {\bf 82}, 1362 (1999)
[arXiv:hep-ph/9809490];\\
F.~Finelli and R.~H.~Brandenberger,
Phys.\ Rev.\ D {\bf 62}, 083502 (2000)
[arXiv:hep-ph/0003172].
%
\bibitem{thermal}
R. Allahverdi, B. A. Campbell and J. Ellis, Nucl. Phys. B {\bf 579}, 355 (2000)
[arXiv:hep-ph/0001122];\\ 
A. Anisimov and M. Dine, Nucl. Phys. B {\bf 619}, 729 (2001)
[arXiv:hep-ph/0008058]. 
%




\end{references}
\end{document}